\documentclass[twocolumn,showpacs,preprintnumbers,amsmath,amssymb]{revtex4}

\usepackage{graphicx}
\usepackage{dcolumn}
\usepackage{bm}

\begin{document}


\title{Metastable Random Field  Ising model with exchange enhancement:
a simple model for Exchange Bias}

\author{Xavier Illa}
\email{xit@ecm.ub.es}
\author{Eduard Vives}
\email{eduard@ecm.ub.es}
\author{Antoni Planes}
\email{toni@ecm.ub.es}
\affiliation{%
Dept. d'Estructura i Constituents de la Mat\`eria, Universitat de Barcelona \\
Diagonal 647, Facultat de F\'{\i}sica, 08028 Barcelona, Catalonia}

\date{\today}

\begin{abstract}
We present a  simple model that allows hysteresis  loops with exchange
bias to be reproduced. The model is a modification of the $T=0$ random
field Ising  model driven  by an external  field and  with synchronous
local relaxation  dynamics.  The main novelty  of the model  is that a
certain fraction  $f$ of  the exchange constants  between neighbouring
spins is enhanced  to a very large value $J_E$.   The model allows the
dependence of the exchange bias and other properties of the hysteresis
loops to be analyzed as a function of the parameters of the model: the
fraction $f$  of enhanced bonds,  the amount of the  enhancement $J_E$
and the amount  of disorder which is controlled  by the width $\sigma$
of the Gaussian distribution of the random fields.
\end{abstract}



\maketitle

\section{Introduction}
\label{Intro}
Hysteresis   and   metastability   are   intriguing   phenomena   with
implications in both fundamental and applied physics. Magnetic systems
are  the  prototypical  example  of thermodynamic  systems  exhibiting
hysteresis cycles for which different theoretical approaches have been
proposed  \cite{Bertotti1998}.  Besides  the  classical micro-magnetic
analysis, based on a continuous description of the magnetic properties
of the system, more recently a lot of efforts have been devoted to the
study  of lattice  models.  For  example, the  zero-temperature Random
Field Ising model  (RFIM) driven by an external  field with convenient
metastable  dynamics   has  been  very   successful  in  qualitatively
explaining some basic properties  of rate independent hysteresis loops
\cite{Sethna1993,Vives1994,Obrado1999,Vives2001}.   The most important
achievement of this model has  been to give a simultaneous explanation
of the effect of disorder on the hysteresis loops and the existence of
Barkhaussen noise  with critical properties.  Less  attention has been
paid to  the use  of such models  for understanding  other interesting
features of the hysteresis  loops such as remanence, coercivity, minor
loop properties or exchange bias (EB) \cite{Nogues1999,Berkowitz1999},
which is the property we will focus our attention on here.

We  present a modification  of the  zero-temperature RFIM  that allows
magnetic  hysteresis  loops  with  EB  to  be  reproduced.   The  main
characteristic  of EB  is that  the hysteresis  loops,  represented as
magnetization $m$  versus external applied field $H$,  are not centred
on $H=0$  but exhibit a  displacement in the  field axis by  an amount
$H_{EB}$ (exchange  bias field) \cite{Nogues1999,Berkowitz1999}.  This
property  has  received  a   lot  of  attention  recently,  since  the
possibility   of  finding   systems   with  large   EB  has   enormous
technological interest  \cite{Dieny1991}.  Experimentally EB  has been
found  in different magnetic  systems \cite{Nogues1999,Berkowitz1999}.
The basic  ingredient for  EB is the  existence of  interfaces between
ferromagnetic (FM) and antiferromagnetic (AFM) systems, where coupling
can be induced after field  cooling from above the Neel temperature of
the AFM.  This heat treatment  freezes some of the magnetic moments at
the interface which are supposed  to be responsible for the occurrence
of  EB.   The prototype  is  a  FM/AFM  bilayer, for  instance  Co/CoO
\cite{Miltenyi2000},      NiFe/NiMn      \cite{Li2000},     Fe/FeF$_2$
\cite{Nogues1996} and Fe/MnF$_2$  \cite{Leighton2000}.  This effect is
also observed  in granular  systems formed by  small particles  with a
ferromagnetic  core covered  by their  native  antiferromagnetic oxide
\cite{Gangopadhyay1993}.

Different  models have  been proposed  to understand  EB.   Although a
basic    qualitative   explanation    was   given    40    years   ago
\cite{Meiklejohn1962} a  deep understanding of the  phenomenon has not
yet  been  achieved  \cite{Kiwi2001,Stamps2000}.   Different  features
remain   unclear:    the   role   played   by    the   AFM   thickness
\cite{Ambrose1996,Miltenyi2000},   the  formation   of   domain  walls
\cite{Malozemoff1987,Mauri1987},  whether the  frozen spins  belong to
the  FM  or  to  the  AFM layer  \cite{Kiwi1999},  etc...   Especially
intriguing is the  fact that EB not only  occurs, in uncompensated AFM
layers which exhibit a net  magnetization after being cooled, but also
in  compensated  AFM  layers  with zero  net  interface  magnetization
\cite{Koon1997}.

The aim of the present paper  is to introduce a very simple model with
a  new mechanism  for the  explanation  of EB  in totally  compensated
layers.  The model  is based on a lattice  spin system with metastable
dynamics for  which some  of the exchange  interactions show  a marked
enhancement.  In section \ref{Model}  the Hamiltonian and the detailed
mestastable dynamics are presented.   In section \ref{Results} we show
the results of the numerical simulations.  In section \ref{Discussion}
we discuss  the possible physical origin of  the exchange enhancement.
In  section \ref{Comparison}  we compare  with  available experimental
data  and,  finally, in  section  \ref{Conclusions}  we summarize  and
conclude.

\section{Model}
\label{Model}
The model is intended to reproduce the properties of the ferromagnetic
layer only, and  regards the AFM part to be  totally quenched and thus
does not contribute to  the net magnetization (compensated AFM layer).
Consequently, we consider the 2d  RFIM on the square lattice, although
a generalization to  bulk ferromagnets or thin layers  could easily be
implemented. Note,  however, that the  AFM plays an indirect  role, as
will be discussed later.

The mathematical formulation of the  model is very similar to the RFIM
on a square lattice with size  $N=L\times L$.  On each lattice site we
define  a  spin  variable  $S_i$  which takes  values  $\pm  1$.   The
Hamiltonian, in reduced units, reads:
\begin{equation}
{\cal H}  = -\sum_{ij}^{nn} J_{ij}  S_i S_j -  \sum_{i}^N h_i S_i  - H
\sum_{i}^N S_i
\label{hamiltonian}
\end{equation}
The first sum is  the ferromagnetic exchange contribution that extends
over nearest  neighbour pairs  ($J_{ij}>0$).  The second  sum accounts
for the  interaction with quenched  random fields $h_i$,  which stands
for the disorder  present in any ferromagnetic system.   The last term
is the  interaction with the  external driving field $H$.   The random
fields $h_i$  are independent and distributed according  to a Gaussian
probability density:
\begin{equation}
p(h_i) = \frac{1}{\sqrt{2 \pi} \sigma} e^{-\frac{h_i^2}{2 \sigma^2}}
\end{equation}
where  $\sigma$ is  the standard  deviation of  the random  fields and
controls the  amount of  disorder in the  system.  The novelty  of the
model is in  the values of the exchange  constants $J_{ij}$, which are
not equal for all spin pairs: we consider that $J_{ij}=J$ except for a
fraction   $f$  of   the  bonds   (selected  at   random)   for  which
$J_{ij}=J_E>>J$.  This  fraction of bonds  is supposed to  contain the
effect of the quenched antiferromagnetic layer.  A physical reason for
this  local   exchange  enhancement  will  be   discussed  in  section
\ref{Discussion}.   Such  a bond  distribution  can be  mathematically
expressed as:
\begin{equation}
p(J_{ij}) = (1-f) \delta(J_{ij}-J) + f \delta(J_{ij}-J_E)
\end{equation}
We  have focused  our  study in  the  region of  small  values of  $f$
($f<0.06$).    The  magnetization   of  the   system  is   defined  as
$m=\sum_{i=1}^{N} S_i /  N$. For the analysis of  the hysteresis loops
we use the  so called synchronous local relaxation  dynamics.  This is
the standard dynamics used in previous studies of the zero-temperature
RFIM \cite{Sethna1993}. Each spin $S_i$ flips according to the sign of
its local field $H_i$ given by:
\begin{equation}
H_i = \sum_{j=1}^4 J_{ij} S_j + H + h_i
\label{localfield}
\end{equation}
where the  first sum  extends over the  four neighbours of  $S_i$.  We
start with a value of $H$ large enough so that the stable situation is
given by  all the spins $S_i=1$.   We decrease the  external field $H$
until  $H_i$ vanishes on  a certain  spin. The  spin is  then reversed
keeping  $H$  constant. This  reversal  may  destabilize  some of  the
neighbouring spins  which are  then reversed simultaneously.   This is
the beginning  of an  avalanche.  The avalanches  proceed until  a new
stable  situation  with  all   the  spins  $S_i$  aligned  with  their
respective  local  fields $H_i$  is  reached.   We  can then  continue
decreasing the external field $H$.

Most of  the calculated  properties are averaged  over a  large number
($\sim  10^3$) of  different realizations  of disorder.   Averages are
indicated by the  symbol $\langle \cdot \rangle $.   We will consider,
without  loss of  generality,  that $J=1$.   Therefore,  from now  on,
magnetic fields and energies are given in units of $J$.
 
\section{Results}
\label{Results}
\begin{center}
\begin{figure}[th]
\includegraphics[width=7cm,clip]{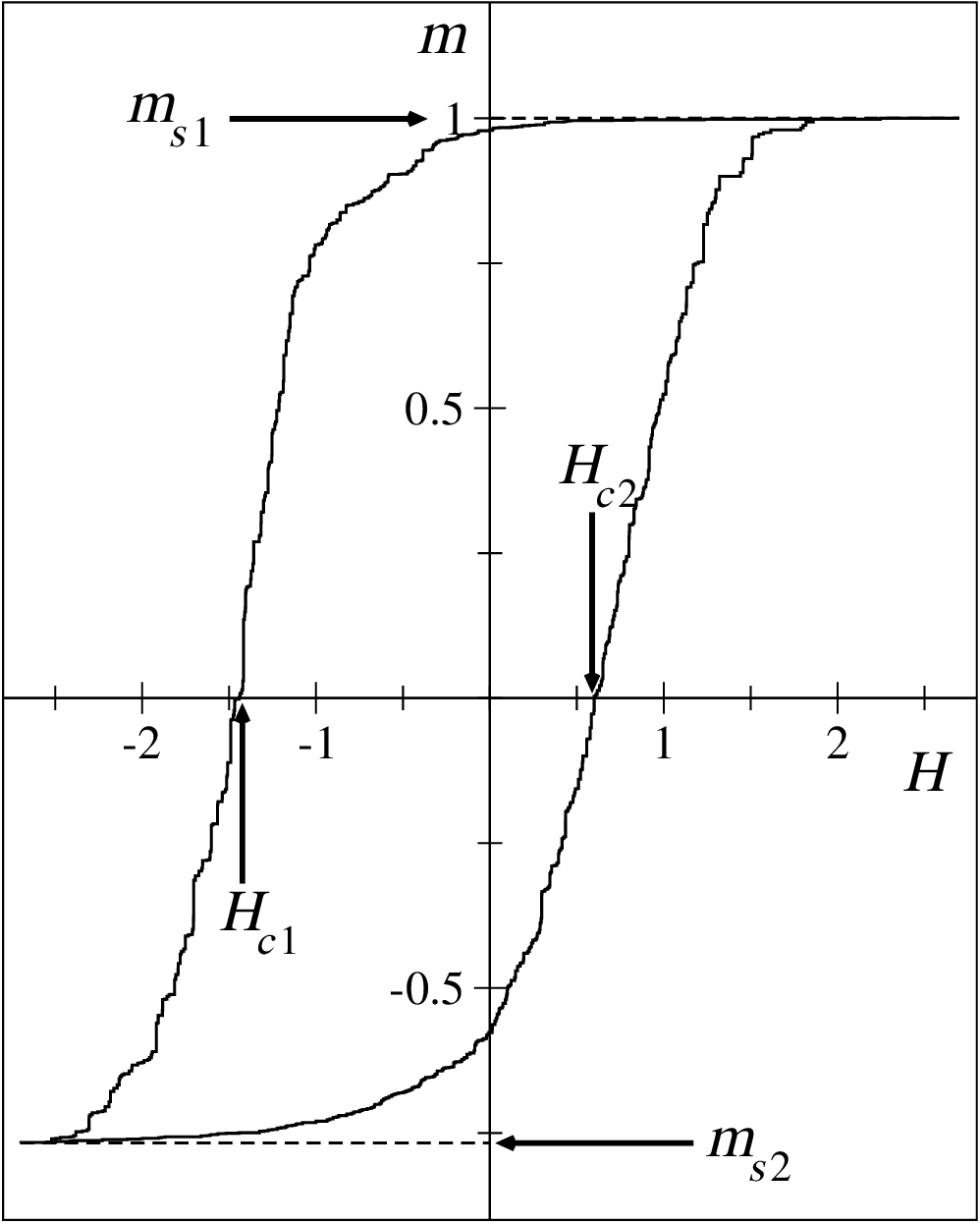}
\caption{\label{fig1} Example of a hysteresis loop exhibiting exchange
bias obtained  from a  numerical simulation of  an $L=50$  system with
$J_E=20$,  $f=0.03$ and  $\sigma=1.65$.  The  external field  has been
swept from 2.7  to -2.7.  $H_{c1}$ and $H_{c2}$  indicate the coercive
fields of the decreasing and the increasing branches respectively.}
\end{figure}
\end{center}

In  Fig.\  \ref{fig1} we  present  an  example  of a  hysteresis  loop
obtained  with  the numerical  simulation  of  a  system with  $L=50$,
$J_E=20$,  $f=0.03$ and  $\sigma=1.65$. The  external field  is cycled
between $H=\pm 2.7$.  As can be seen, the loop exhibits remarkable EB.

\begin{center}
\begin{figure}[th]
\includegraphics[width=7.5cm,clip]{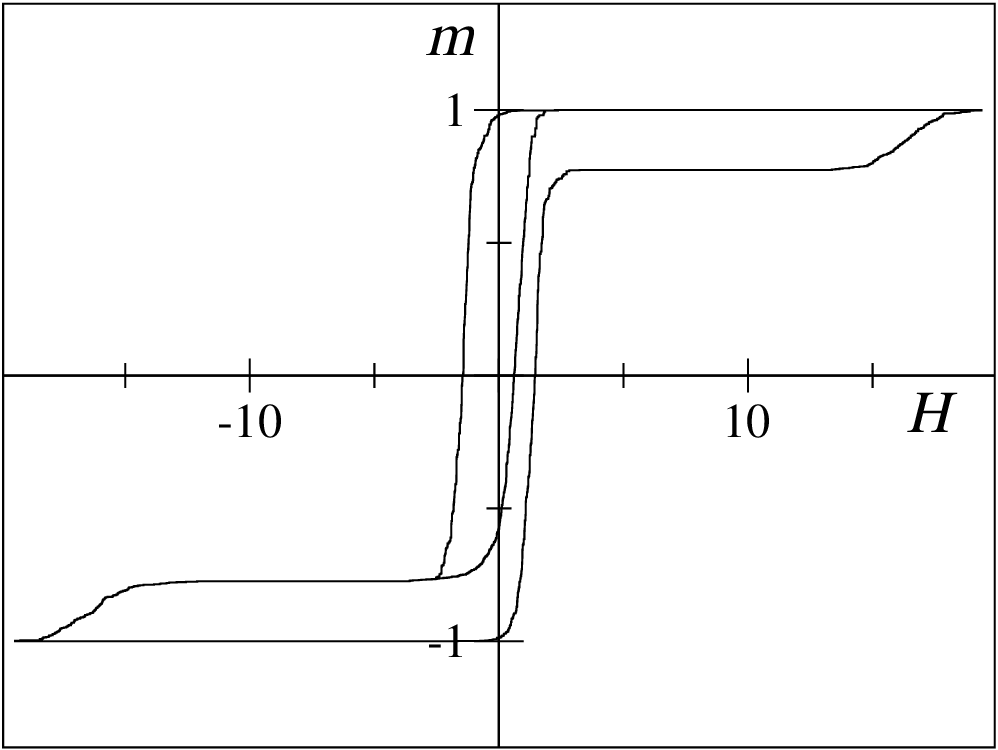}
\caption{\label{fig2}  The  same  example  of Fig.\  1  revealing  the
partial loop with exchange bias  and the total loop between $H=\pm 19.4$
which is symmetric.}
\end{figure}
\end{center}

At first glance  it may look surprising that the  model defined in the
previous  section displays  such asymmetry,  since the  Hamiltonian is
totally  symmetric  under  the  changes  $S_i  \rightarrow  -S_i$  and
$H\rightarrow -H$.   The reason is  that the hysteresis loop  shown in
figure  \ref{fig1} corresponds,  strictly speaking,  to a  minor loop.
This  is revealed  in  Fig.\  \ref{fig2} where  the  system is  cycled
between $H=\pm 19.4$ (which is a  field that is one order of magnitude
larger than the  coercive field $H_{c1}$).  Due to  the existence of a
tiny  fraction of  very large  exchange interactions,  the  total loop
exhibits long,  flat plateaux in  which the system  behaves reversibly
exactly as if  it was saturated.  Only when  cycling between extremely
large  values of  the external  field, does  one obtain  the symmetric
hysteresis loop. Therefore, the loops with EB are incomplete loops and
are accompanied by a magnetization shift.

In  order  to  perform  a  quantitative analysis  of  this  behaviour,
averaging over different realizations  of disorder is carried out. The
hysteresis  loops   are  systematically  obtained   according  to  the
following protocol: decreasing the field from $H=+\infty$ to $2 H_{c1}
< 0$ and increasing the  field again to $H=+\infty$, where $H_{c1}$ is
the  coercive field  in the  decreasing branch.   We also  compute the
pseudo coercive  field $H_{c2}$ in  the increasing branch  (see figure
\ref{fig1}).

The criteria for choosing the value $2 H_{c1}$ as a returning point is
similar to  the criteria used  in many experimental cases.   One could
easily change  this limit to  $3H_{c1}$ or $4H_{c1}$  without changing
the results provided  that $J_E$ is large enough.   This can be easily
understood from  the flat tails in  the full hysteresis  loop shown in
figure \ref{fig2}.

\begin{center}
\begin{figure}[th]
\includegraphics[width=7.5cm,clip]{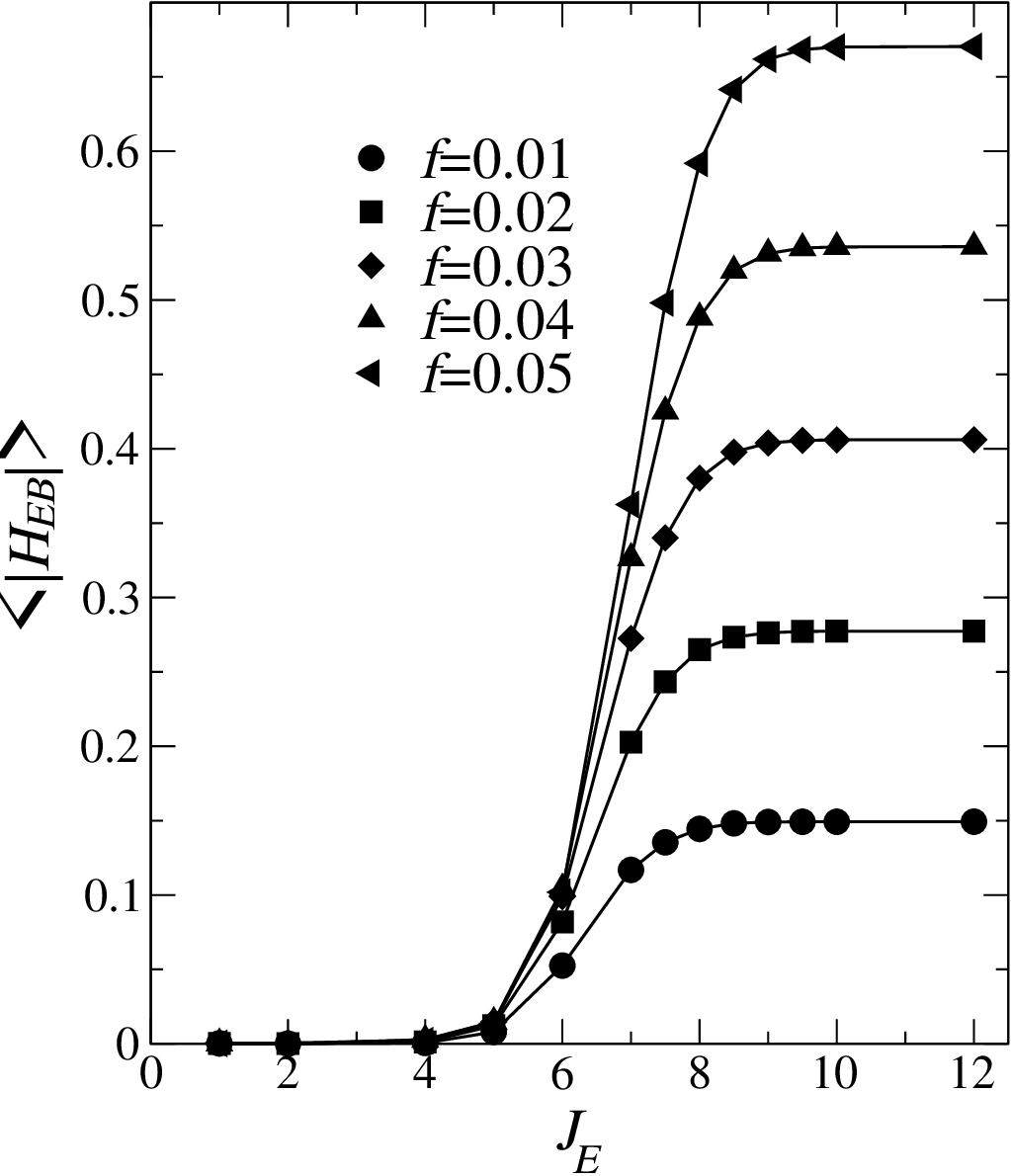}
\caption{\label{fig3}  Dependence  of  the  exchange bias  $\langle  |
H_{EB}| \rangle $ on the exchange enhancement $J_E$ for $\sigma=1$ and
different values  of $f$  as indicated by  the legend. Data  have been
obtained  by   averaging  1000  realizations  of  a   system  of  size
$L=50$. The lines are a guide  to the eye.  Statistical error bars are
smaller than the symbol size.}
\end{figure}
\end{center}

\begin{center}
\begin{figure}[th]
\includegraphics[width=7.5cm,clip]{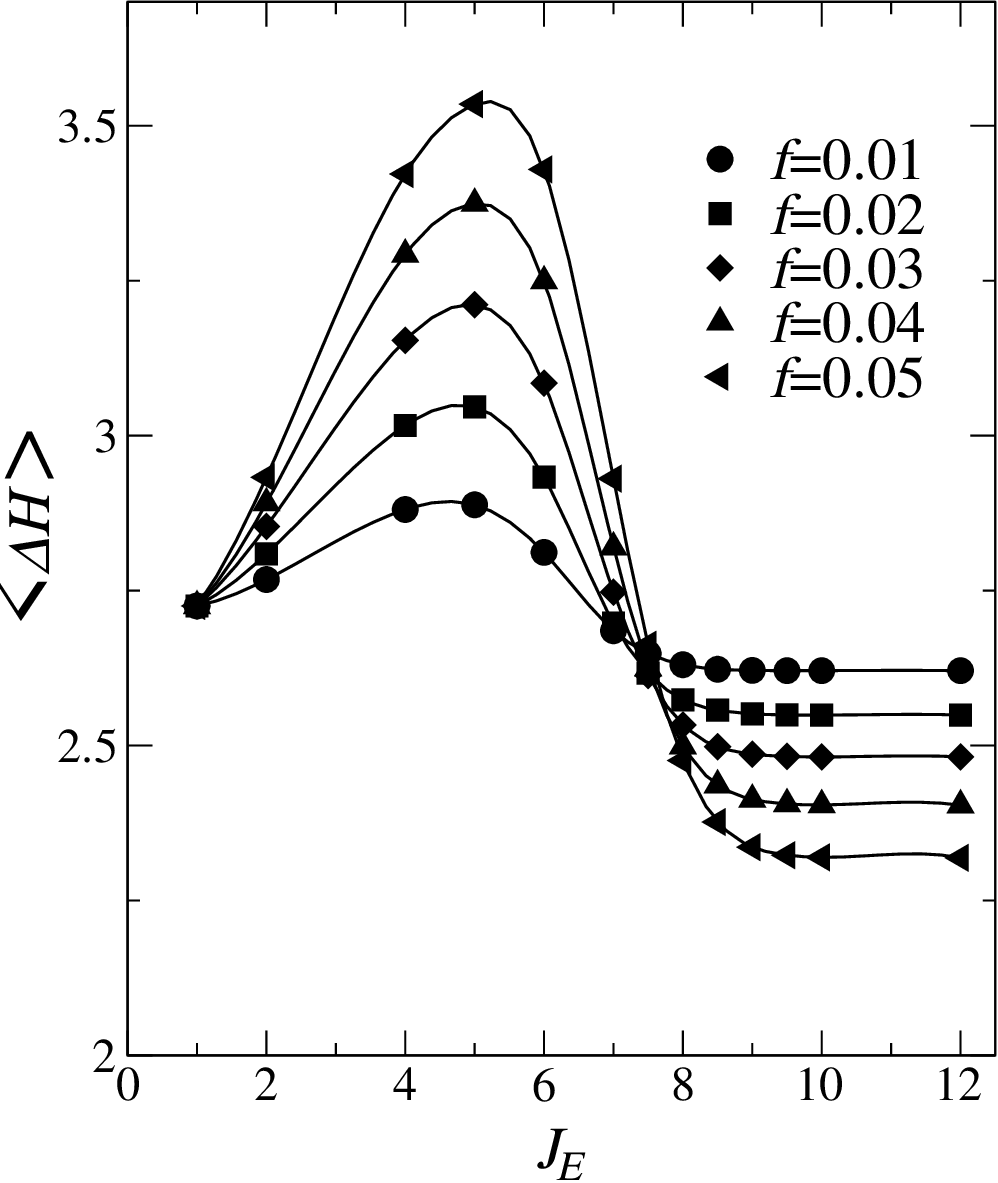}
\caption{\label{fig4} Dependence  of the coercivity  $\langle \Delta H
\rangle  $  on  the  exchange  enhancement $J_E$  for  $\sigma=1$  and
different values  of $f$ as indicated  by the legend.   Data have been
obtained  by   averaging  1000  realizations  of  a   system  of  size
$L=50$. The lines are a guide  to the eye.  Statistical error bars are
smaller than the symbol size.}
\end{figure}
\end{center}

The EB field $H_{EB}$, the coercivity $\Delta H$ and the magnetization
shift $m_B$ are defined as:
\begin{equation}
H_{EB}= \frac{H_{c1}+H_{c2}}{2} 
\end{equation}
\begin{equation}
\Delta H = H_{c2}-H_{c1}
\end{equation}
\begin{equation}
m_B = \frac{m_{s1}+m_{s2}}{2}=\frac{1+m_{s2}}{2}
\end{equation}
According to this definition, the loops shifted to the left on the $H$
axis  (as  occurs with  the  loops in  the  present  paper) will  have
negative exchange bias field.   Figures \ref{fig3} and \ref{fig4} show
the dependence of $\langle | H_{EB} | \rangle$ and $\Delta H$ on $J_E$
for $\sigma=1$ and different values of $f$ as indicated by the legend.
Even  for  very  low values  of  $f$,  $\langle  | H_{EB}  |  \rangle$
increases and saturates for large enough values of $J_E$.

In  the case  of coercivity  (Fig. \ref{fig4})  two  important results
should be emphasized: the increase  (almost 40 \% in certain cases) in
coercivity for  intermediate values of  $J_E$ and the saturation  at a
constant value (which depends on  $f$) for large $J_E$.  Such limiting
values at large $J_E$ however,  are smaller than the coercivity of the
system without exchange enhancement.

\begin{center}
\begin{figure}[th]
\includegraphics[width=7.5cm,clip]{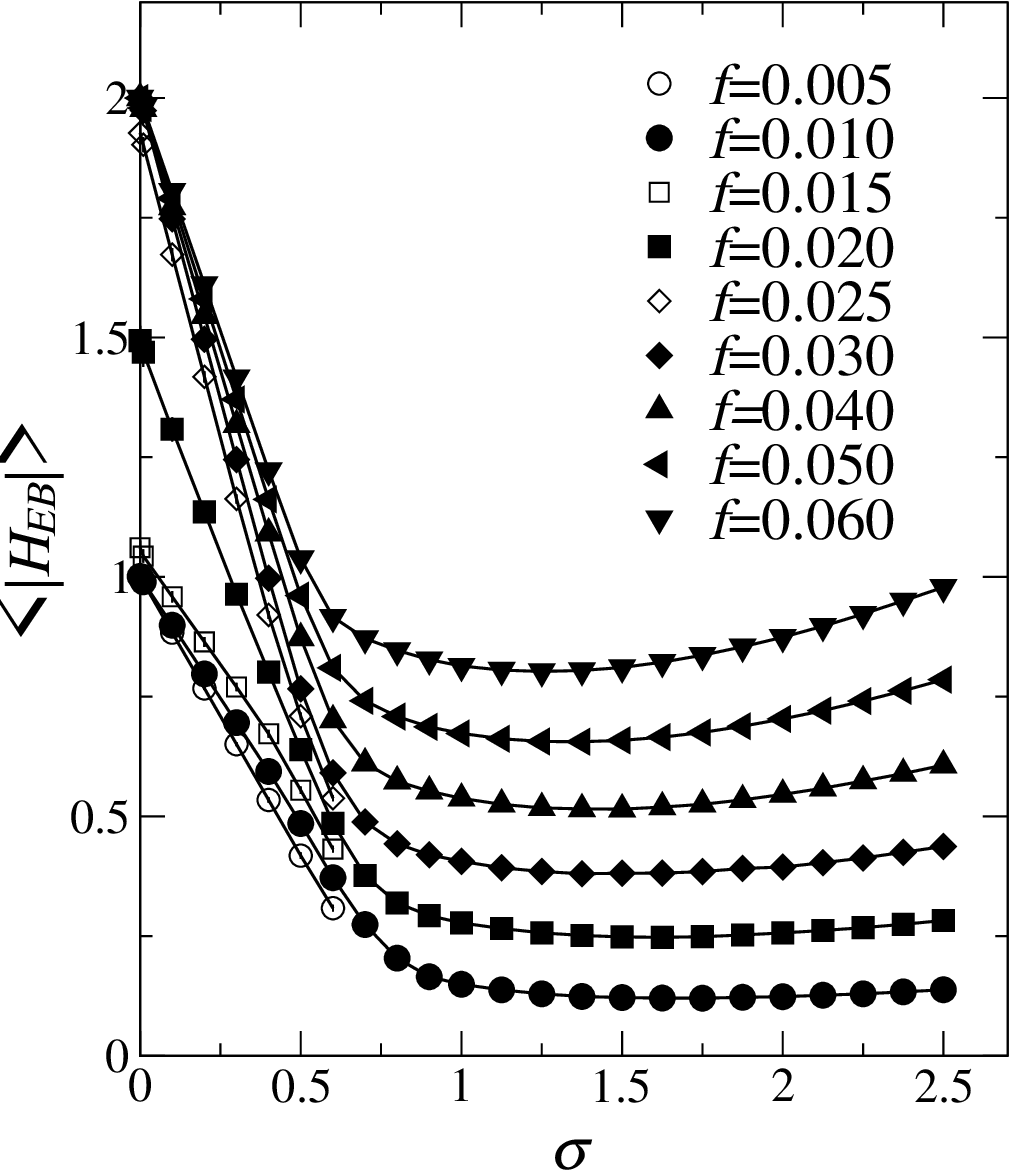}
\caption{\label{fig5}  Dependence   of  the  exchange   bias  $\langle
|H_{EB}| \rangle $ on the amount of disorder $\sigma$ for $J_E=20$ and
different values  of $f$  as indicated by  the legend. Data  have been
obtained  by   averaging  1000  realizations  of  a   system  of  size
$L=50$. The lines  are a guide to the eye.  Statistical error bars are
smaller than the symbol size.}
\end{figure}
\end{center}

\begin{center}
\begin{figure}[th]
\includegraphics[width=7.5cm,clip]{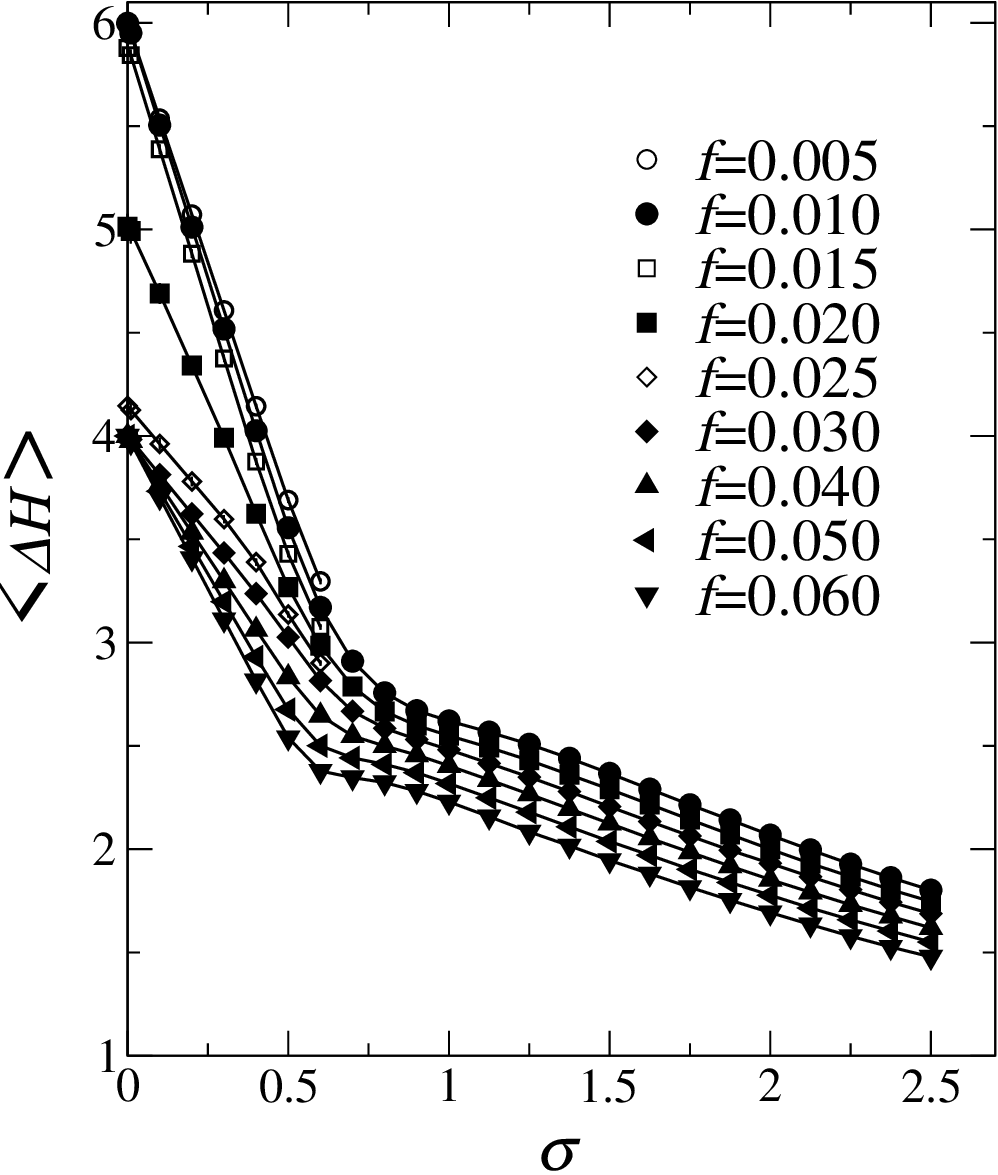}
\caption{\label{fig6} Dependence  of the coercivity  $\langle \Delta H
\rangle  $  on  the  amount  of disorder  $\sigma$  for  $J_E=20$  and
different values  of $f$ as indicated  by the legend.   Data have been
obtained  by   averaging  1000  realizations  of  a   system  of  size
$L=50$. The lines are a guide  to the eye.  Statistical error bars are
smaller than the symbol size.}
\end{figure}
\end{center}

In order  to analyze  the dependence of  the system properties  on the
amount of disorder $\sigma$, we choose  a value of $J_E$ that is large
enough so that  $\langle | H_{EB} | \rangle$  has reached the limiting
maximum  value   (see  Fig.\  \ref{fig3}).    Figures  \ref{fig5}  and
\ref{fig6}  show the  behaviour of  $\langle |  H_{EB} |  \rangle$ and
$\langle \Delta H \rangle $ versus $\sigma$ for $J_E=20$ and different
values of  $f$ as indicated  by the legend.  Unexpectedly,  $\langle |
H_{EB} |  \rangle$ shows non-monotonic behaviour  with $\sigma$, first
decreasing until a minimum is  reached, but which increases slowly for
large amounts of disorder.

The  marked variation  of $\langle  | H_{EB}  | \rangle$  and $\langle
\Delta  H \rangle$  for values  of $f$  between 0.015  and  0.025 when
$\sigma \rightarrow 0$ is associated  with the fact that the ascending
part  of the hysteresis  loop is  very sensitive  to the  existence of
nuclei of unreversed spins in  the negative magnetized state. For very
low values  of $f$ we expect  that all the nuclei  of unreversed spins
will be formed  by two positive spins joined by  an enhanced bond. The
negative spins  surrounding such a  nucleus will flip (in  the $\sigma
\rightarrow  0$  limit)  around  $H=2$  (see  Eq.   \ref{localfield}).
However,  for larger  values of  $f$, larger  nuclei will  exist.  For
instance, a nucleus formed by  three spins joined by two perpendicular
bonds, both with exchange enhancement, acts as a nucleating seed which
triggers the avalanches towards  the positive magnetization phase when
$H=0$.  When $f$  is large enough, such that  the probability for such
nuclei is  significantly different from  zero, the coercive  field for
the ascending branch  decreases from 2 to 0,  thus increasing $H_{EB}$
and decreasing $\Delta H$.

The  non-monotonic behaviour  of  the EB  when  disorder is  increased
better seen  by plotting $\langle |  H_B | \rangle /  \langle \Delta H
\rangle$, which is a dimensionless  quantity and is more relevant from
the experimental point of view.  Note that for large $f$ and $\sigma$,
we can  find strongly biased  hysteresis loops for which  $H_{c1}$ and
$H_{c2}$  are negative.  An  example, obtained  by sweeping  the field
between $H=\pm 4.5$ is shown in figure \ref{fig8}.
\begin{center}
\begin{figure}[th]
\includegraphics[width=7.5cm,clip]{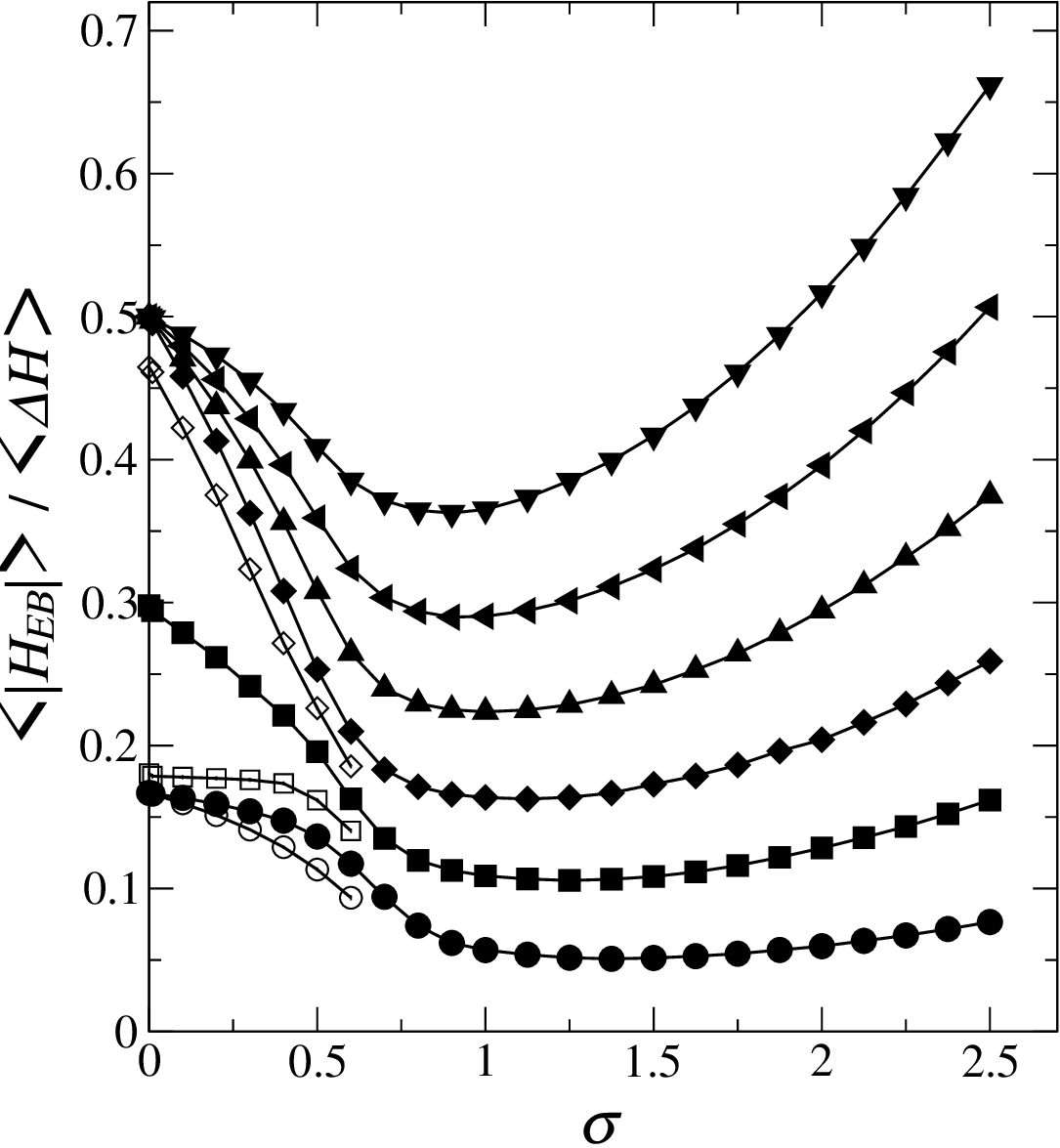}
\caption{\label{fig7}  Dependence   of  the  relative   exchange  bias
$\langle |H_{EB} |\rangle /\langle \Delta  H \rangle$ on the amount of
disorder  $\sigma$  for  $J_E=20$  and  different  values  of  $f$  as
indicated by the legend in Fig. \ref{fig6}. Data have been obtained by
averaging 1000 realizations of a  system of size $L=50$. The lines are
a guide to the eye. Statistical error bars are smaller than the symbol
size.}
\end{figure}
\end{center}
\begin{center}
\begin{figure}[th]
\includegraphics[width=7.5cm,clip]{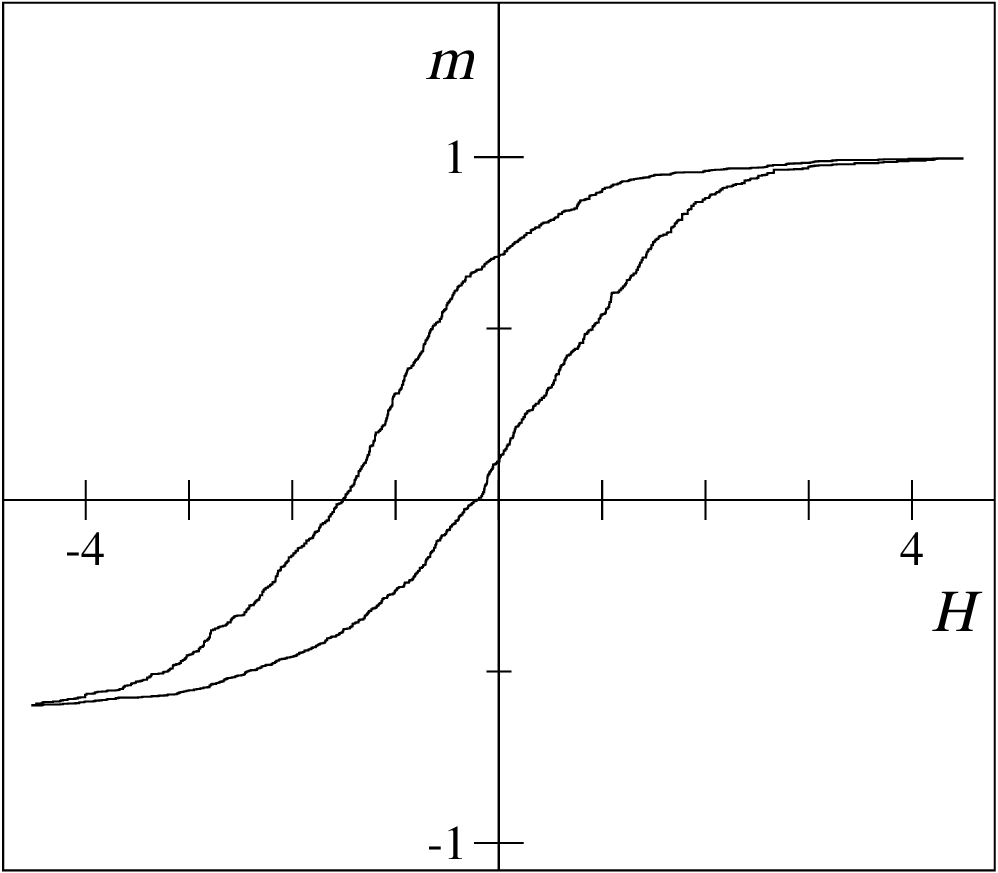}
\caption{\label{fig8}   Hysteresis  loop   exhibiting   exchange  bias
obtained from a numerical simulation of a $L=50$ system with $J_E=20$,
$f=0.05$ and  $\sigma=3$. The external  field has been  cycled between
$H=\pm 4.5$.}
\end{figure}
\end{center}

\section{Discussion}
\label{Discussion}
\begin{center}
\begin{figure}[th]
\includegraphics[width=7.5cm,clip]{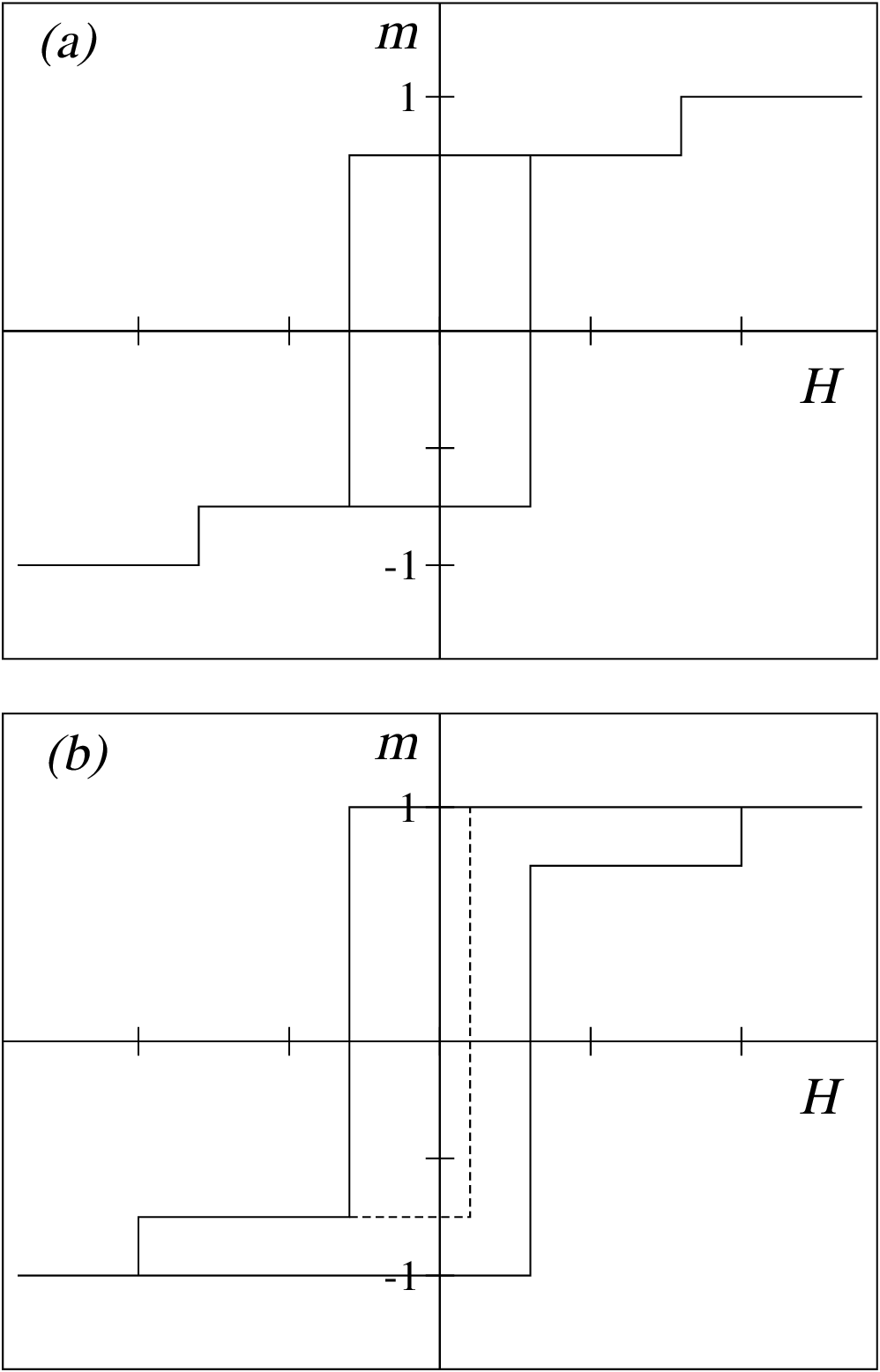}
\caption{\label{fig9} Schematic  examples  of   the   hysteresis loops
obtained for a model with strongly enhanced $\pm  h$ random fields and
for  strong enhanced random  bonds.  In  both cases  the full loop  is
symmetric, but the minor loop only exhibits EB for the enhanced random
bonds.} 
\end{figure}
\end{center}

One of the reasons for the present work was to obtain hysteresis loops
with EB by modifying the zero-temperature RFIM  as little as possible.
The   straightforward, naive idea  would be  to consider non-symmetric
distributions of  random  fields, for  which the  average $\langle h_i
\rangle \neq  0$ will create the  displacement of the  loop.  From our
point of view, this will correspond to the  effect of an uncompensated
AFM layer.   Constraining ourselves  to  the inclusion  of compensated
disorder,  we  have found that   any  symmetric distribution of random
fields  cannot  give EB.  To understand  this,  suppose that a certain
fraction of spins is pinned  by very positive and (symmetrically) very
negative random   fields.   Fig.\ \ref{fig9}(a)   shows the  schematic
hysteresis loop  corresponding to such a  system.  The spins with more
positive  random  fields, which   are the  last   to  reverse   in the
decreasing branch, will be the first  to flip in the increasing branch
of the full hysteresis   loop.   Therefore, the full hysteresis   loop
(symmetric, without EB) will overlap with  the minor hysteresis loop,
as shown in    Fig.\    \ref{fig9}a.  In  contrast,   when the    bond
distribution is distorted, as done in the model presented in this work
(and which preserves the symmetry properties of the Hamiltonian in Eq.
\ref{hamiltonian}), the  spins with larger ferromagnetic coupling will
be the last to reverse in  the decreasing branch  and also the last to
be  reversed in  the increasing branch.   Thus,  minor loops  will not
coincide with  the  full loops and  will easily   exhibit large EB  as
indicated in Fig.\ \ref{fig9}b. 

The  justification of  the  proposed model  would  require a  physical
explanation for  the exchange enhancement phenomena.   Recently it has
been  suggested  that spin-waves  in  the  FM/AFM  interface could  be
responsible   for   an    enhancement   of   the   exchange   coupling
\cite{Suhl1998}.  Here  we propose a different mechanism  based on the
existence of quenched disorder in  the AFM layer. The most common case
is the existence of antiphase domain boundaries as considered in Refs.
\onlinecite{Miltenyi2000},                      \onlinecite{Nowak2002},
\onlinecite{Keller2002}  and \onlinecite{Misra2001}.   One  can assume
that the exchange  interaction between the magnetic moments  in the FM
layer has two contributions: the first comming from the direct overlap
of the  electronic wave functions of the  atoms in the FM  layer and a
second arising  from a  superexchange interaction through  the overlap
with  the electrons  in  the  AFM layer.   The  existence of  quenched
disorder in the  AFM layer can modify this  second contribution of the
exchange interaction  giving rise to the  exchange enhancement.  Since
the energies associated  with the broken AFM bonds  can be higher than
the FM exchange energies (for  instance due to the existence of strong
anisotropy), it  is plausible to imagine  that the defects  in the AFM
can influence the FM exchange interactions.

\begin{center}
\begin{figure}[th]
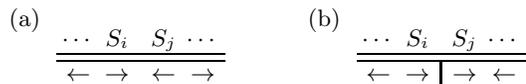

(a)
\raisebox{-0.5cm}{
\begin{tabular}{ccccc} 
$\cdots$  & $S_i$ & &$S_j$ & $\cdots$  \\ \hline\hline
$\leftarrow$ & $\rightarrow$ & & $\leftarrow$ & $\rightarrow$ \\
\end{tabular}}
\hspace{1cm}(b)
\raisebox{-0.5cm}{
\begin{tabular}{ccccc} 
$\cdots$  & $S_i$ & &$S_j$ & $\cdots$  \\  \hline\hline 
$\leftarrow$ & $\rightarrow$ &\vline & $\rightarrow$ & $\leftarrow$ \\
\end{tabular}}
\caption{\label{figtab}  Schematic representation  of two  spins $S_i$
and $S_j$ of  the FM layer on (a)  an ordered AFM layer and  (b) on an
antiphase domain boundary.}
\end{figure}
\end{center}

We would like  to give a  possible mathematical formulation for such a
physical  mechanism within the  framework of  lattice models.  Let  us
consider that the interaction between two neighbouring spins $S_i$ and
$S_j$ of the FM layer is given by: 
\begin{equation}
E_{ij} = - J_0 S_i S_j - K S_i S_j \sigma_i \sigma_j
\end{equation}
where $\sigma_i$ and $\sigma_j$  are the spin variables describing the
magnetic moments in the AFM layer that sit exactly below the $S_i$ and
$S_j$ spins  of the FM layer. The   constant $J_0>0$ accounts  for the
exchange interactions in the free FM  layer and $K>0$ accounts for the
coupling between the two layers. This coupling term can include, in an
effective form,  the interactions with many atomic  layers in the AFM.
Figure  \ref{figtab} shows a schematic representation  of the spins of
the FM  and the AFM  layer in  two situations: (a)  corresponds to the
normal situation for an ordered  AFM layer, whereas (b) corresponds to
the case in which the AFM exhibits an antiphase domain boundary. 

In case (a), the energy of the $i,j$ pair is $E_{ij}=-(J_0-K)S_i S_j$,
whereas in case (b) (or on any antiphase boundary) the energy is given
by $E_{ij}=-(J_0+ K)  S_i S_j$.  The  ratio  between the two  exchange
constants is $J_E=  (J_0+ K)/(J_0-K)$, which can  be  much larger than
$1$ when $K$ is close but smaller than $J_0$. 

\section{Comparison with experiment}
\label{Comparison}
The idea that  exchange bias could be a minor  loop effect was already
suggested  to  explain  the  influence  of the  AFM  spin-flop  in  EB
\cite{Nogues2000}. Moreover, it has been shown experimentally that the
uncompensated spins  which have been  suggested to result  in exchange
bias \cite{Takano1997} and which could result in the proposed exchange
enhancement,    can    be    reversed    at   high    enough    fields
\cite{Ambrose1996} (much higher than  those usually used in exchange
bias studies).  Hence,  in this case EB could also  be considered as a
minor loop effect to a certain extent.

Moreover, the  peak observed in  the behaviour of the  coercivity with
$J_E$ (see fig. \ref{fig4})  could be correlated with the experimental
observation  of  peaks in  the  coercive  field $H_c(t_{AFM})$  (where
$t_{AFM}$ is the thickness of the AFM layer) close to the critical AFM
thickness for  the onset of  EB or in  $H_c(T)$ close to  the N{\'e}el
temperature (see  figures 10 and 13  in Ref. \onlinecite{Nogues1999}).
One  could consider  that  as  the AFM  thickness  or the  temperature
increases, the FM-AFM coupling, which causes the enhancement of J$_E$,
decreases  and  is  consequently  the  origin of  the  observed  $H_c$
increase,  in  agreement  with  theory.   Similarly,  the  changes  in
$\langle  H_{EB}\rangle$ and  $\langle  \Delta H  \rangle$ with  J$_E$
could also be correlated with the behaviour experimentally observed in
Fe/MnF$_2$ under a cooling field \cite{Leighton2000}.

Another interesting result  of this model is the  behaviour of EB with
the amount  of  disorder $\sigma$, as   seen in Figs.\  \ref{fig5} and
\ref{fig7}.  Experimentally  the role played by disorder  in EB is not
clear.    In  some cases  there is  evidence  that increasing disorder
increases   $H_{EB}$, whereas in other cases   the opposite effect has
been found  \cite{Nogues1999b}.  Our  model  is  able to  explain both
possibilities.   Moreover,   it   is  noteworthy  that    experimental
non-monotonic dependence   of  $H_{EB}$ on disorder   (e.g. roughness,
irradiation damage  or structural disorder)   have also been  reported
\cite{Miltenyi2000,Leighton1999,Mougin2000}. 

\begin{center}
\begin{figure}[th]
\includegraphics[width=7.5cm,clip]{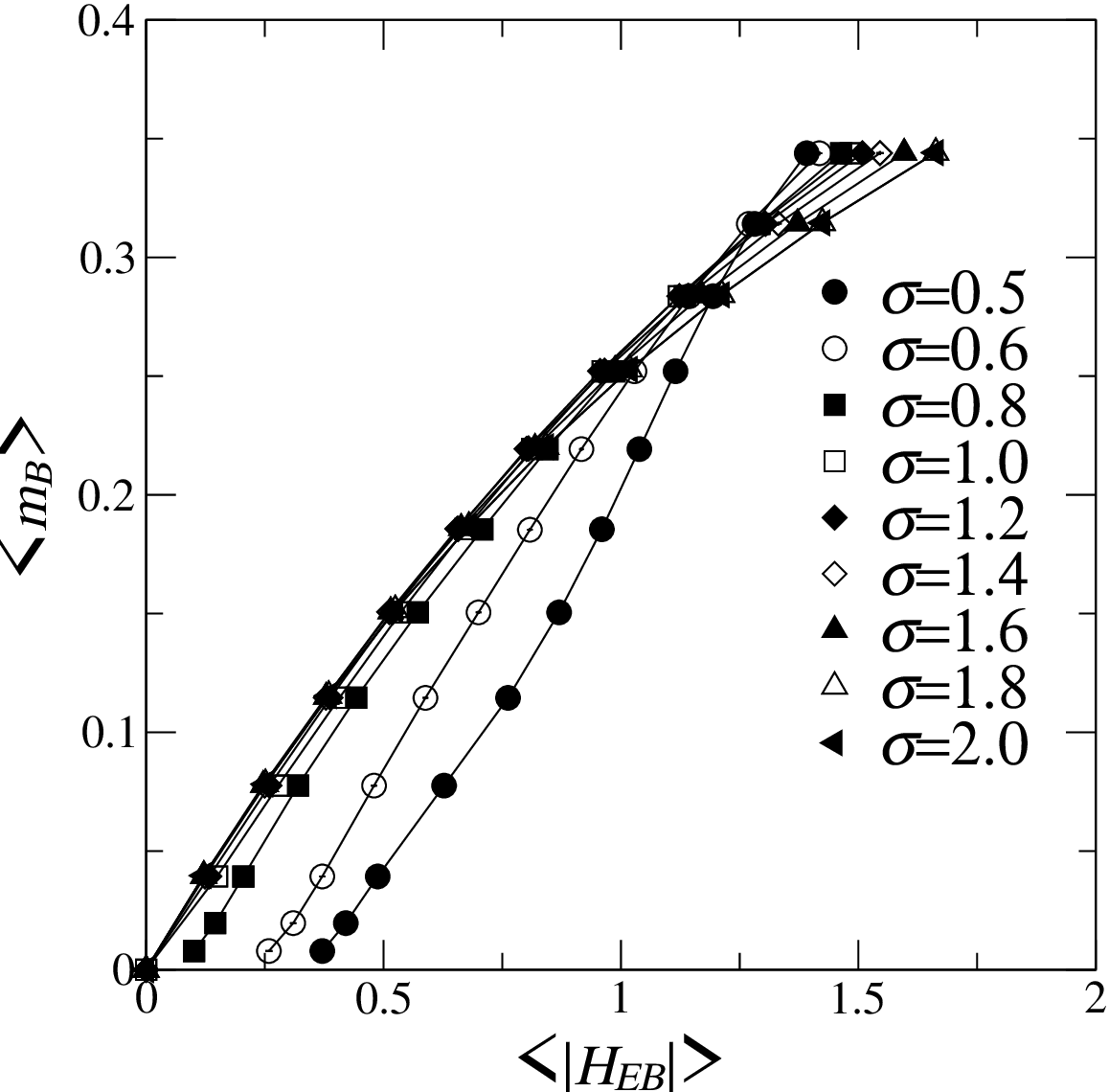}
\caption{\label{fig10} Magnetization  shift $m_B$ versus  $H_{EB}$ for
different  values of  $f$  as  indicated by  the  legend.  The  points
correspond to systems with  different values of $\sigma$, ranging from
$0$ to $2$. The lines are a guide to the eye. Statistical error bars are
smaller than the symbol size.}
\end{figure}
\end{center}

Finally, another  remarkable result from  our model is that  the loops
exhibiting EB also exhibit a vertical shift in the magnetization axis.
This      effect       has      been      observed      experimentally
\cite{Meiklejohn1957,Radu2002,Nogues1999b,Nogues2000b,Keller2002}.
However, since the  fraction $f$ is very small, the  shift can also be
very  small  and  in  some   cases,  would  be  difficult  to  observe
experimentally.   Fig.\  \ref{fig10}   shows  the  dependence  of  the
magnetization shift  $m_{B}$ on $H_{EB}$ for different  values of $f$.
The points correspond to different values of $\sigma$ ranging from $0$
to $2$. In  view of these results, it would  be interesting to measure
such displacements in different experimental systems.

\section{Summary  and conclusions}
\label{Conclusions}
We have  presented a simple model for  the study of Exchange Bias. The
model is based on the RFIM driven by an external field with metastable
dynamics.  The key ingredient  is the existence  of a tiny fraction of
ferromagnetic bonds, which are  strongly enhanced. This creates a long
reversible   plateaux    in  the    decreasing   branch.   When   this
pseudo-saturated state  is reached,  the  reversal of the field  gives
rise to an  EB loop.  Different  properties of  these loops  have been
computed  as a  function of  the fraction of  enhanced   bonds and the
amount of disorder   in  the system.  We  have  suggested   a possible
physical  mechanism to  justify  the  existence  of such  an  exchange
enhancement   in a FM   layer  on a  AFM  layer with  antiphase domain
boundaries,  based on the existence  of superexchange coupling between
the two layers. 

The main conclusions of the paper are: (i)  EB is due  to a minor loop
effect; (ii) compensated AFM layers can  exhibit exchange biased loops
with   a  concomitant magnetization   shift;  (iii)  many experimental
phenomena related  to exchange  bias such  as:  peaks in the  coercive
field,  magnetization   shifts,     marked coercivity    increase   or
non-monotonic  dependence of $H_{EB}$ on  disorder can successfully be
reproduced with this model. 

\section{Acknowledgements}
We acknowledge  many fruitful  discussions with J.Nogu\'es.  This work
has   received   financial  support   from   CICyT  (Spain),   project
MAT2001-3251  and  CIRIT  (Catalonia), project  2000SGR00025.   X.Illa
acknowledges financial support from DGICyT.

\end{document}